\documentclass[%
 reprint,
 amsmath,amssymb,
 aps,
]{revtex4-2}

\usepackage{graphicx}
\usepackage{dcolumn}
\usepackage{xcolor}
\usepackage{bm}
\usepackage{hyperref}
\usepackage{slashed}
\usepackage{subfloat}
\usepackage{pgfplots,subfigure}

\begin{document}

\preprint{APS/123-QED}

\title{\textcolor{black}{An innovative model for coupled fermion-antifermion pairs}}

\author{Abdullah Guvendi}
\email{abdullah.guvendi@erzurum.edu.tr}
\affiliation{Department of Basic Sciences, Erzurum Technical University, 25050, Erzurum, Türkiye}

\author{Omar Mustafa}
\email{omar.mustafa@emu.edu.tr}
\affiliation{Department of Physics, Eastern Mediterranean University, G. Magusa, north Cyprus, Mersin 10 - Türkiye}

\date{\today}

\begin{abstract}
\textcolor{black}{Understanding the behavior of fermion-antifermion (\(f\overline{f}\)) pairs is crucial in modern physics. These systems, governed by fundamental forces, exhibit complex interactions essential for particle physics, high-energy physics, nuclear physics, and solid-state physics. This study introduces a novel theoretical model using the many-body Dirac equation for \(f\overline{f}\) pairs with an effective position-dependent mass (i.e., \(m \rightarrow m + \mathcal{S}(r)\)) under the influence of an external magnetic field. To validate our model, we show that by modifying the mass with a Coulomb-like potential, \(m(r) = m - \alpha/r\), where \(-\alpha/r\) is the Lorentz scalar potential \(\mathcal{S}(r)\), our results match the well-established energy eigenvalues for \(f\overline{f}\) pairs interacting through the Coulomb potential, without approximation. By applying adjustments based on the Cornell potential (i.e., \(\mathcal{S}(r) = kr - \alpha/r\)), we derive a closed-form energy expression. We believe this unique model offers significant insights into the dynamics of \(f\overline{f}\) pairs under various interaction potentials, with potential applications in particle physics. Additionally, it could be extended to various \(f\overline{f}\) systems, such as positronium, relativistic Landau levels for neutral mesons, excitons in monolayer transition metal dichalcogenides, and Weyl pairs in monolayer graphene sheets.}

\end{abstract}

\maketitle


\section{Introduction}

\textcolor{black}{The study of $f\overline{f}$ systems has been a subject of profound interest in modern physics, driven by the quest to understand the fundamental constituents of matter and the underlying forces that govern their behavior. Fermions and antifermions are fundamental particles that obey the Pauli exclusion principle, meaning no two fermions can occupy the same quantum state simultaneously. Fermions include particles such as electrons, quarks, protons, and neutrons, while their corresponding antiparticles are positrons, antiquarks, antiprotons, and antineutrons \cite{p-phys}. These particles are essential building blocks of matter, with fermions forming standard matter and antifermions representing their antimatter counterparts. When fermions and antifermions meet, they can annihilate each other, releasing energy in the form of photons or other particles \cite{p-phys}. Their interactions, governed by the fundamental forces, play a crucial role in many areas of physics, including particle physics, high-energy physics, and astrophysics, providing deep insights into the nature of the universe and the fundamental forces that govern it.}

\textcolor{black}{\(f\overline{f}\) pairs, such as positronium \cite{p-laser,positronium}, quarkonium \cite{quarkonium0}, exciton \cite{exciton,new-1,new-2}, and Weyl pair \cite{graphene}, serve as critical systems for exploring fundamental interactions and symmetries in various domains of physics. Positronium \cite{p-laser,positronium}, an electron-positron bound state, provides a unique laboratory for precision tests of quantum electrodynamics (QED) and annihilation processes. Quarkonium \cite{quarkonium0}, consisting of a quark-antiquark ($q\overline{q}$) pair, is instrumental in probing the strong force described by quantum chromodynamics (QCD) \cite{qq0}, with its energy levels and transitions offering insights into confinement and hadronization. Excitons \cite{exciton,new-1,new-2}, formed by electron-hole pairs in semiconductors, play a vital role in understanding and harnessing optoelectronic properties of materials, impacting the development of advanced photovoltaic and light-emitting devices. Weyl pairs in topological materials like graphene manifest exotic relativistic behavior, leading to phenomena such as the chiral anomaly and providing a platform for studying relativistic quantum mechanics in condensed matter systems \cite{graphene}. These \(f\overline{f}\) pairs not only deepen our comprehension of fundamental physics but also drive technological advancements across multiple disciplines.}

\textcolor{black}{The dynamics of $f\overline{f}$ pairs can be studied using various types of interaction potentials. In the context of the Dirac equation, the electromagnetic minimal potential and the Lorentz scalar potential represent two different ways of incorporating external fields into the relativistic wave equation of a particle \cite{g-plb1,g-landau,PDM-dirac}. The electromagnetic minimal potential involves coupling the Dirac field to an electromagnetic field through the four-vector potential \(\mathcal{A}_\mu = (\phi, \mathbf{A})\), where \(\phi\) is the scalar potential and \(\textbf{A}\) is the vector potential. In a flat spacetime, the Dirac equation in the presence of an electromagnetic field is modified by replacing the partial derivative \(\partial_\mu\) with the covariant derivative \(D_\mu = \partial_\mu + ie\mathcal{A}_\mu\) \cite{g-plb1,g-landau}, where \(e\) is the electric charge of the particle. This modification ensures gauge invariance and describes the interaction between charged particles and the electromagnetic field, as seen in QED. Conversely, the Lorentz scalar potential, involves modifying the mass term of the Dirac equation. Instead of coupling through the derivative, the scalar potential \(\mathcal{S}(\textbf{x})\) is added directly to the mass term, modifying it to \(m \rightarrow m + \mathcal{S}(\textbf{x})\) \cite{PDM-dirac}. This type of potential does not affect the gauge properties of the Dirac field but alters the effective mass of the particle locally. The primary difference lies in their interaction with the Dirac field: the electromagnetic minimal potential changes the kinetic term by modifying the derivatives to include the vector potential, reflecting the coupling to an electromagnetic field, while the Lorentz scalar potential modifies the mass term, affecting the intrinsic energy of the particle directly.}

The study of the dynamics governing $f\overline{f}$ systems, within the context of the well-known electromagnetic minimal interaction potentials and Lorentz scalar potentials, necessitates examining developed many-body equations. These equations typically involve formulating the free Dirac Hamiltonian for each $f$ and $\overline{f}$, alongside incorporating interaction potentials. However, a significant challenge arises from relative time issues, stemming from retardation effects, which are often overlooked in these phenomenological equations despite their intrinsic involvement in a multi-temporal scenario. Further complications arise in selecting appropriate interaction potentials and determining the overall angular momentum of the systems formed by fermions and antifermions within these equations. Traditionally, interaction potentials are simplified to either a single boson or photon exchange potential. Nevertheless, these phenomenologically established many-body equations struggle to maintain complete covariance in curved spacetimes \cite{breit,bs}.

On the other hand, non-relativistic quantum mechanics employs one-time equations comprising individual particle free Hamiltonians and an interaction term, with wave functions depending on the spatial coordinates of each particle. The introduction of the Dirac equation spurred research towards formulating a comprehensive two-body Dirac equation, notably initiated by Breit \cite{breit}, which incorporated two free Dirac Hamiltonians and a modified interaction potential resembling the Darwin potential in electrodynamics. However, challenges arise under conditions of long-range interactions or high particle velocities due to retardation effects. Subsequent efforts by Bethe and Salpeter \cite{bs}, utilizing quantum field theory, encountered relative time issues, necessitating approximations like instantaneous interactions to address interacting particles. Decades later, Barut proposed a fully covariant many-body Dirac equation encompassing spin algebra and the most general electric and magnetic potentials \cite{barut}. Barut's equation, characterized by a spin algebra involving Kronecker products of Dirac matrices, results in a $16\times16$ dimensional matrix equation in $(3+1)$-dimensions. Achieving separation between angular and radial components involves the utilization of group theoretical techniques \cite{nuri}. Nonetheless, solving the resulting set of $16$ radial equations remains a formidable challenge, even for familiar systems such as one-electron atoms. This challenge arises from the interdependence of radial equations, resulting in pairs of coupled second-order wave equations-a hurdle that demands meticulous attention \cite{nuri}. However, recent years have indicated that the Barut equation is entirely solvable when addressing low-dimensional systems or systems demonstrating specific dynamical symmetries, irrespective of whether they inhabit flat or curved spaces \cite{guvendi-c1,g-btz,g-plb2,g-epjc,omar-4,td3}. This equation has made significant strides in its application, marking notable breakthroughs. 

\textcolor{black}{In this methodological proposal, our investigation aims to elucidate the behavior of \(f\overline{f}\) pairs influenced by well-known minimal coupling and Lorentz scalar potentials in 2+1 dimensions, using Barut's fully covariant many-body Dirac equation derived from quantum electrodynamics through the action principle \cite{barut}}. Our study is organized as follows. In section \ref{sec2}, we introduce our new model for a general $f\overline{f}$ pair subject to a Lorentz scalar potential in (2+1)-dimensional spacetime with polar coordinates \((t, r, \phi)\). We apply an external magnetic field using the $3$-vector potential, introduced through \(\mathcal{A}^{f}_{\phi}=\frac{\mathcal{B}_{0}}{2\hbar c} r^{2}_{f}\) and \(\mathcal{A}^{\overline{f}}_{\phi}=\frac{\mathcal{B}_{0}}{2\hbar c} r^{2}_{\overline{f}}\), where $\mathcal{B}_{0}$ represents the external magnetic field. We consider an effective position-dependent mass setting for the Cornell-type potential, given by \(m(r)=2m-\alpha/r+kr\) \cite{Cornell}. The corresponding covariant two-body Dirac equation is then reduced to a two-dimensional one-body radial \textcolor{black}{Schrödinger-like equation}. In section \ref{sec3.1}, we solve the model without the external magnetic field (i.e., $\mathcal{B}_{0} = 0$) and obtain an exact closed-form analytical solution for \(m(r)=2m-\alpha/r\) . We thoroughly discuss our exact solution and compare it with available data. This is, nevertheless, complemented by a brief discussion, in section \ref{sec3.1.1}, on the implications of our results for excitons in monolayer transition metal dichalcogenides (TMDCs). Additionally,  in section \ref{sec3.2}, we provide a conditionally exact solution for the case with \(m(r)=2m-\alpha/r+kr\) in the presence of an external magnetic field ($\mathcal{B}_{0} \neq 0$). Finally, we summarize and discuss our findings in section \ref{sec4}. 

\section{\label{sec2}{A unique model}} \label{sec2}

\textcolor{black}{In this section, we introduce a model for a general \(f\overline{f}\) pair subject to a Lorentz scalar potential under the influence of an external magnetic field in (2+1)-dimensional spacetime with polar coordinates \((t, r, \phi)\)}. With negative signature, the metric describing the spacetime background is written as the following: \(ds^2 = c^2dt^2 - dr^2 - r^2d\phi^2\), where $c$ is the speed of light. Under such spacetime settings, the covariant many-body Dirac equation reads \cite{g-btz,g-plb2,g-epjc}:
\begin{equation}
\begin{split}
&\left\lbrace \mathcal{H}^{f} \otimes \gamma^{t^{\overline{f}}}+ \gamma^{t^{f}}\otimes \mathcal{H}^{\overline{f}} \right\rbrace \Psi(x_{\mu}^{f},x_{\mu}^{\overline{f}})=0,\\
&\mathcal{H}^{f(\overline{f})}= \slashed{\nabla}_{\mu}^{f(\overline{f})} +i\tilde{m}(x_{\mu}^{f},x_{\mu}^{\overline{f}}) \mathcal{I}_{2},\\
&\slashed{\nabla}_{\mu}^{f (\overline{f})}=\gamma^{\mu^{f (\overline{f})}}\left(\partial_{\mu}^{f (\overline{f})}+i\frac{e_{f(\overline{f})}\mathcal{A}^{f(\overline{f})}_{\mu}}{\hbar c}-\Gamma_{\mu}^{f (\overline{f})} \right).\label{eq1}
\end{split}
\end{equation}
Here, $\tilde{m}$ is defined as $mc/\hbar$, with $m$ indicating the rest mass of individual quarks, $e$ is the electric charge, $\mathcal{A}_{\mu}$ is the $3$-vector potential, $\hbar$ is the reduced Planck constant, Greek indices indicates coordinates within the spacetime ($x^{\mu}=t,r,\phi$), and $\Psi(x_{\mu}^{f},x_{\mu}^{\overline{f}})$ stands for the bi-local spinor that relies on the spacetime position vectors ($x_{\mu}^{f},x_{\mu}^{\overline{f}}$) of the particles. Additionally, $\mathcal{I}_2$ denotes $2\times2$ unit matrix, fermions (antifermions) are referred to as $f$ ($\overline{f}$). Within equation (\ref{eq1}), the space-dependent Dirac matrices $\gamma^{\mu}$ can be obtained through the relationship: $\gamma^{\mu}=e^{\mu}_{(a)}\gamma^{(a)}$ \cite{g-npb}, where $e^{\mu}_{(a)}$ represents the inverse tetrad fields and $\gamma^{(a)}$ symbolizes the space-independent Dirac matrices. These matrices, referred to as $\gamma^{(a)}$, are expressed in terms of the Pauli matrices $(\sigma_{x}, \sigma_{y}, \sigma_{z})$ 
\begin{equation}
\gamma^{(0)}=\sigma_{z},\quad \gamma^{(1)}=i\,\sigma_{x},\quad \gamma^{(2)}=i\,\sigma_{y},\label{FDM}
\end{equation}
where $i$ symbolizes the imaginary unit ($i=\sqrt{-1}$). In the realm of a (2+1)-dimensional metric with a negative signature, where the Minkowski tensor ($\eta_{(a)(b)}$) takes the form $\eta_{(a)(b)}=\textrm{diag}(+,-,-)$, the determination of inverse tetrad fields follows the expression: $e^{\mu}_{(a)}=g^{\mu\tau}e{\tau}^{(b)}\eta_{(a)(b)}$. Here, the contravariant metric tensor is denoted as $g^{\mu\tau}$, while $e_{\tau}^{(b)}$ represents the tetrad fields derived from $g_{\mu\tau}=e_{\mu}^{(a)}e_{\tau}^{(b)}\eta_{(a)(b)}$, where $g_{\mu\tau}$ stands for the covariant metric tensor. Moreover, the spinorial affine connections $\Gamma_{\mu}$ in Eq. (\ref{eq1}) can be determined by using $\Gamma_{\lambda}=\frac{1}{4}\left[e^{(a)}_{\nu_{,\lambda}}e^{\tau}{(a)}-\Gamma_{\nu\lambda}^{\tau} \right]\frac{1}{2}\left[\gamma^{\mu},\gamma^{\nu}\right]$, with $\Gamma_{\nu\lambda}^{\tau}$ representing the Christoffel symbols obtained through $\Gamma_{\nu \lambda}^{\tau} = \frac{1}{2}g^{\tau \epsilon}\left[\partial_{\nu} g_{\lambda \epsilon} + \partial_{\lambda} g_{\nu \epsilon} - \partial_{\epsilon} g_{\nu \lambda} \right]$ \cite{g-npb}. Hence, we can uncover the relevant generalized Dirac matrices ($\gamma^{\mu}$) and the spinorial affine connections' non-zero elements ($\Gamma_{\mu}$) for Dirac fields as the following:
\begin{equation}
\begin{split}
&\gamma^{t^{f(\overline{f})}}=\gamma^{(0)}/c,\quad \gamma^{r^{f(\overline{f})}}=\gamma^{(1)},\\
&\gamma^{\phi^{f(\overline{f})}}=\frac{\gamma^{(2)}}{r^{f(\overline{f})}},\quad \Gamma^{f(\overline{f})}_{\phi}=\frac{i}{2}\sigma_{z}.\label{GDM-SC}
\end{split}
\end{equation}
In the affine spin connections $\Gamma^{f(\overline{f})}_{\mu}$, only one component stands out amidst the zeros. Consequently, we have
\begin{equation}
\gamma^{\phi^{f(\overline{f})}}\Gamma^{f(\overline{f})}_{\phi}=-\frac{i}{2r^{f(\overline{f})}}\sigma_{x}.
\end{equation}
It is known that the external magnetic field can be introduced through the angular component of the $3$-vector potential, namely \(\mathcal{A}^{f}_{\phi}=\frac{\mathcal{B}_{0}}{2\hbar c} r^{2}_{f}\) and \(\mathcal{A}^{\overline{f}}_{\phi}=\frac{\mathcal{B}_{0}}{2\hbar c} r^{2}_{\overline{f}}\) \cite{g-landau,g-plb2}, where $\mathcal{B}_{0}$ is the external magnetic field. To enhance clarity, let us explicitly introduce the corresponding two-body Dirac equation as $\hat{M}\Psi=0$, where $\hat{M}$ is
\begin{equation}
\begin{split}
&\gamma^{t^{f}}\otimes\gamma^{t^{\overline{f}}}\left[\partial_{t}^{f}+\partial_{t}^{\overline{f}} \right]+\gamma^{r^{f}}\partial_{r}^{f}\otimes \gamma^{t^{\overline{f}}}+ \gamma^{t^{f}}\otimes \gamma^{r^{\overline{f}}}\partial_{r}^{\overline{f}}\\
&+\gamma^{\phi^{f}} \otimes\gamma^{t^{\overline{f}}}\slashed{\partial}_{\phi}^{f}+\gamma^{t^{f}}\otimes \gamma^{\phi^{\overline{f}}}\slashed{\partial}_{\phi}^{\overline{f}}\\
&+i\tilde{m}(x_{\mu}^{f},x_{\mu}^{\overline{f}})\left[\mathcal{I}_{2}\otimes \gamma^{t^{\overline{f}}}+\gamma^{t^{f}}\otimes \mathcal{I}_{2}\right]\\
&-\left[\gamma^{\phi^{f}}\Gamma_{\phi}^{f}\otimes \gamma^{t^{\overline{f}}}+\gamma^{t^{f}}\otimes \gamma^{\phi^{\overline{f}}}\Gamma_{\phi}^{\overline{f}} \right],\label{eq4}
\end{split}
\end{equation}
where 
\begin{equation*}
\begin{split}
\slashed{\partial}_{\phi}^{f}\rightarrow \partial_{\phi}^{f}+i\frac{e\mathcal{B}_{0}}{2\hbar c}r^{2}_{f},\quad \slashed{\partial}_{\phi}^{\overline{f}}\rightarrow \partial_{\phi}^{\overline{f}}-i\frac{e\mathcal{B}_{0}}{2\hbar c}r^{2}_{\overline{f}},
\end{split}
\end{equation*}
since \(e_{\overline{f}}=-e_{f}\) \cite{g-landau}. Upon thorough examination of the spacetime interval at hand, we can express the spacetime-dependent bi-spinor $\Psi(t,r,R)$ in a factorized manner, enabling us to decompose $\Psi$ into $\textcolor{black}{e^{-i\,\frac{\mathcal{E}}{\hbar}\, t}e^{i\vec{K} \cdot \vec{R}}\tilde{\Psi}(\vec{r})}$, where $\tilde{\Psi}(\vec{r})=e^{i s \phi}(\psi_{1}(r)\,\psi_{2}(r)\,\psi_{3}(r)\,\psi_{4}(r))^{T}$. Here, \textcolor{black}{$\mathcal{E}$ denotes the relativistic energy}, while $\vec{r}$ and $\vec{R}$ denote the spatial position vectors for the relative motion and center of mass motion, respectively. $\vec{K}$ represents the center of mass momentum vector, $s$ is the total spin of the $f\overline{f}$ pair, and $^{T}$ signifies the transpose of the $r$-dependent spinor. Following the standard approach for analyzing two-body systems, we introduce the relative motion and center of mass motion coordinates as follows \cite{guvendi-c1}:
\begin{equation*}
\begin{split}
&R_{x^{\mu}}=\frac{x^{\mu^{f}}}{2}+\frac{x^{\mu^{\overline{f}}}}{2},\quad r_{x^{\mu}}=x^{\mu^{f}}-x^{\mu^{\overline{f}}},\\
&x^{\mu^{f}}=\frac{1}{2}r_{x^{\mu}}+R_{x^{\mu}},\quad x^{\mu^{\overline{f}}}=-\frac{1}{2}r_{x^{\mu}}+R_{x^{\mu}},\\
&\partial_{x_{\mu}}^{f}=\partial_{r_{x^{\mu}}}+\frac{1}{2}\partial_{R_{x^{\mu}}},\quad \partial_{x_{\mu}}^{\overline{f}}=-\partial_{r_{x^{\mu}}}+\frac{1}{2}\partial_{R_{x^{\mu}}}, \label{eq4-}
\end{split}
\end{equation*}
for a $f\overline{f}$ pair. It is interesting to observe that the combination $\partial_{x^{\mu}}^{f}+\partial_{x^{\mu}}^{\overline{f}}$ can be simplified to $\partial_{R_{x^\mu}}$. Essentially, this means that the system's evolution, which is influenced by the \textcolor{black}{relativistic energy $\mathcal{E}$}, is intertwined with proper time, symbolized by $\partial_{R_{t}}$. If we assume that the center of mass remains stationary at the spatial origin, we can derive a set of equations that govern the relative motion of the pair in the center of mass frame, where the total momentum ($\vec{K}$) of the system is zero. 

By adding and subtracting these equations, we can derive the following equation set (two of which are algebraic) by assuming the total mass of the system depends only on the relative radial distance ($r$) between the particles
\begin{equation}
\begin{split}
&\textcolor{black}{\tilde{\mathcal{E}}} \chi_{1}(r)-2\tilde{m}(r)\chi_{2}(r)+2\hat{\Lambda}\chi_{4}(r)-\frac{4\left(s+\mathcal{B}r^2\right)}{r}\chi_{3}(r)=0,\\
&\textcolor{black}{\tilde{\mathcal{E}}} \chi_{2}(r)-2\tilde{m}(r)\chi_{1}(r)=0,\\
&\textcolor{black}{\tilde{\mathcal{E}}} \chi_{3}(r)-\frac{4\left(s+\mathcal{B}r^2\right)}{r}\chi_{1}(r)=0,\\
&\textcolor{black}{\tilde{\mathcal{E}}} \chi_{4}(r)-2\hat{\Lambda} \chi_{1}(r)=0,\label{eqset}
\end{split}
\end{equation}
where $\hat{\Lambda}=\left(\frac{1}{r}+\partial_{r}\right)$, $\textcolor{black}{\tilde{\mathcal{E}}}=\frac{\mathcal{E}}{\hbar c}$, $\mathcal{B}=\frac{e\mathcal{B}_{0}}{8\hbar c}$, and
\begin{equation}
\begin{split}
&\chi_{1}(r)=\psi_{1}(r)+\psi_{4}(r),\quad \chi_{2}(r)=\psi_{1}(r)-\psi_{4}(r),\\
&\chi_{3}(r)=\psi_{2}(r)+\psi_{3}(r),\quad \chi_{4}(r)=\psi_{2}(r)-\psi_{3}(r).
\end{split}
\end{equation} 
These equations yield the subsequent wave equation
\begin{equation}
\partial^{2}_{r}\chi_{1}+\frac{2}{r}\partial_{r}\chi_{1}+\left[\frac{\textcolor{black}{\tilde{\mathcal{E}}}^2-m^{2}_{*}(r)}{4}-\frac{4\left(s+\mathcal{B}r^2\right)^{2}}{r^2}\right]\chi_{1}=0,\label{WE}
\end{equation}
where $m_{*}(r)=2\tilde{m}+\textcolor{black}{\mathcal{S}(r)}$. We emphasize that this equation can be analyzed by making appropriate mass modifications tailored to the systems under study, as demonstrated in the following sections. Therefore, we believe that our model can establish a robust foundation for numerous future studies. We can now begin demonstrating the functionality of our model.

\section{Analytical solutions}\label{sec3}

In this section, we seek analytical solutions to Eq. (\ref{WE}) under two distinct scenarios: i) when \(\textcolor{black}{\mathcal{S}(r)}=-\alpha/r\) and \(\mathcal{B}_{0}=0\) , and then ii) when \(\textcolor{black}{\mathcal{S}(r)}=-\alpha/r+kr\) and \(\mathcal{B}_{0}\neq0\).

\subsection{Analytical solutions for the case where \(\textcolor{black}{\mathcal{S}(r)} = -\alpha / r\) and \(\mathcal{B}_{0} = 0\)}\label{sec3.1}

\vspace{12pt}

\begin{table}[b]
\caption{\label{tab:table1}
A report on binding energy $\mathcal{E}^{b}_{n}$ of singlet positronium. Here, $m_{e}$ is the usual electron/positron mass $\sim 9.10938356 10^{-31}$ [kg], $c=299792458$ [m/s] and $\alpha\sim 1/137$.}
\vspace{7pt}
\begin{ruledtabular}
\begin{tabular}{lcdr}
\multicolumn{1}{c}{\textrm{quantum state}}&
\textrm{$\mathcal{E}^{b}_{n}$ [eV]}\\
\colrule
n=0 & - 6.806 \\
\colrule
n=1 & - 1.701 \\
\colrule
n=2 & - 0.756 \\
\colrule
n=3 & - 0.425 \\
\colrule
n=4 & - 0.272 \\
\end{tabular}
\end{ruledtabular}
\end{table}

If $\textcolor{black}{\mathcal{S}(r)}=-\alpha/r$, where $\alpha$ represents the fine structure constant (approximately $\alpha\sim 1/137$), Eq. (\ref{WE}) can be reformulated using a new dimensionless variable: $\xi=\sqrt{(2\tilde{m})^2-\textcolor{black}{\tilde{\mathcal{E}}}^2}\, r$. This transformation simplifies the equation, and by introducing the ansatz $\chi_{1}(\xi)=\frac{1}{\xi}\chi(\xi)$, we can further reduce it into a more recognizable form
\begin{equation}
\chi^{''}(\xi)+\left[\frac{\tilde{\mu}}{\xi}-\frac{1}{4}+\frac{\frac{1}{4}-\tilde{\nu}^2}{\xi^2}\right]\chi(\xi)=0,\label{Whit}
\end{equation}
where
\begin{equation*}
\tilde{\mu}=\frac{2\tilde{m}\alpha}{2\sqrt{(2\tilde{m})^2-\textcolor{black}{\tilde{\mathcal{E}}}^2}},\quad \tilde{\nu}=\frac{\sqrt{1+\alpha^2+16s^2}}{2}.
\end{equation*}
Solution function of the Eq. (\ref{Whit}) can be expressed in terms of the Confluent Hypergeometric function, $\chi(\xi)=e^{-\frac{\xi}{2}}\xi^{\frac{1}{2}+\tilde{\nu}}\  _1F_{1}(\frac{1}{2}+\tilde{\nu}-\tilde{\mu}, 1+2\tilde{\nu}; \xi)$ around the regular singular point $\xi=0$ \cite{g-btz}. In this context, it is crucial to emphasize that the analyzed system forms a bound pair. To ensure the finiteness and square integrability of $\chi(\xi)$, we need to truncate the Confluent Hypergeometric series to a polynomial of order $n_r\geq 0$. This necessitates satisfying the condition $\frac{1}{2}+\tilde{\nu}-\tilde{\mu}=-n_r$  \cite{g-btz}. Consequently, we derive the quantization condition (\textcolor{black}{$\mathcal{E} \rightarrow \mathcal{E}_{ns}$}) for the formation of such a pair, yielding the following expression:
\begin{equation*}
  \textcolor{black}{\mathcal{E}_{ns}=\pm 2m c^2} \sqrt{1-\frac{\alpha^2}{(2n_r+\sqrt{1+\alpha^2+16s^2}+1)^2}}.  
\end{equation*}
When $s=0$, the relativistic energy spectrum can be rewritten using the power series expansion method, which is typical for one-electron systems
\begin{widetext}
\begin{equation}
\mathcal{E}_{n}\approx 2mc^2 \left\lbrace 1-\frac{\alpha^2}{8\,(n+1)^2}+\frac{(8n+7)\,\alpha^4}{128\,(n+1)^4}-\frac{(16n^3+72n^2+88n+33)\,\alpha^6}{1024\,(n+1)^6} \right\rbrace .\label{eq7}
\end{equation}
\end{widetext}

\begin{table*}
\caption{\label{tab:table2} A report on the binding energy $\mathcal{E}^{b}_{n}$ [eV] of exciton in monolayer TMDCs. Here, $m_{e} \sim 9.10938356 10^{-31}$ [kg], $c=299792458$ [m/s] and $\alpha\sim 1/137$.}
\vspace{7pt}
\begin{ruledtabular}
\begin{tabular}{cccccccccc}
quantum state &$\epsilon_{eff}=3$&$\epsilon_{eff}=4$&$\epsilon_{eff}=5$
&$\epsilon_{eff}=6$&$\epsilon_{eff}=7$&$\epsilon_{eff}=8$&$\epsilon_{eff}=9$&$\epsilon_{eff}=10$\\ \hline
 n=0&-0.756 (see \cite{new-2})  &-0.425  &-0.272 (see \cite{new-1})  &-0.189 & -0.138  & -0.106  & -0.084  & -0.068 \\ \hline
 n=1&-0.189  &-0.106  &-0.068  &-0.047 & -0.034  & -0.026  & -0.021  & -0.017 \\ \hline
n=2&-0.084  &-0.047  &-0.030  &-0.021 & -0.015  & -0.011  & -0.009  & -0.007 \\
\end{tabular}
\end{ruledtabular}
\end{table*}

In this relativistic energy spectrum, the first term represents the total rest mass energy ($2mc^2$), while the term proportional to $\alpha^2$ corresponds to the well-known non-relativistic binding energy. For para-positronium (the singlet spin state of a bound electron-positron pair) in ground state, this binding energy is $\mathcal{E}^{b}_{0} \approx -m_{e}c^2\alpha^2/4 \sim -6.8$ eV \cite{book1,p-laser,positronium} when $m_{e}$ is the usual electron mass. The higher order terms $(\alpha^4, \alpha^6)$ include relativistic corrections. Accordingly, the overall energy conveyed by the annihilation photons (para-positronium decays predominantly into two photons due to the conservation of overall charge conjugation parity) can be calculated, as $\sim 2m_{e}c^2-|\mathcal{E}^{b}|$ \cite{positronium}, where $m_{e}$ is the usual electron/positron mass, since the binding energy is negative. The results from our unique model, as shown in Table \ref{tab:table1}, align with the previously reported binding energy values in \cite{p-laser}.

\subsubsection{A brief discussion on excitons in monolayer TMDCs}\label{sec3.1.1}

This energy spectrum (\ref{eq7}) can be applied, without lose of generality, to an exciton in TMDCs only through $\alpha \rightarrow \alpha/\epsilon_{eff}$, where $\epsilon_{eff}$ is the effective dielectric constant of the surrounding medium \cite{exciton}. These results indicate that the ground state binding energy of the exciton can range from several eV to a few meV when $\epsilon_{eff}$ varies from 2 to 10 (see also Table \ref{tab:table2}). Our results provide insights that can clarify the discrepancies between previously reported observations \cite{new-1,new-2}, but they exclude thermal effects. Furthermore, our findings suggest that the evolution of excitons in a monolayer medium can be regulated by adjusting the effective dielectric constant, which is influenced by the substrate material. It is important to highlight that controlling exciton evolution in monolayer materials can yield numerous advantages. By manipulating exciton properties, such as their binding energies, we can potentially enhance the efficiency of light absorption and emission in these materials. This improvement can lead to the development of more efficient photodetectors, light-emitting devices, and photovoltaic cells. Additionally, monolayer materials hold significant promise for quantum technologies, including quantum dots and quantum information processing. Precise control over exciton evolution is crucial for the creation and manipulation of quantum states, positioning these materials as strong candidates for quantum computing and communication. The ability to regulate exciton properties enables the creation of tunable optoelectronic devices, allowing for the adjustment of optical and electronic characteristics to meet specific requirements. Understanding how excitons behave under magnetic fields can lead to the design of better optoelectronic devices, such as photodetectors, light-emitting diodes (LEDs), and solar cells. Magnetic fields can be used to tune the optical and electronic properties of these devices for optimal performance. TMDCs have unique spin and valley degrees of freedom, which can be manipulated using magnetic fields. Accurate results on excitonic behavior help in developing spintronic and valleytronic devices, where information is carried by spin or valley states rather than charge. Magnetic fields can induce new excitonic states (magneto-excitons) with distinct properties. Studying these states can lead to the discovery of novel physical phenomena and phases of matter. In monolayer TMDCs, strong magnetic fields can give rise to quantum Hall effects \cite{he1,he2}, where excitonic properties play a crucial role. Understanding these effects is important for solid-state physics and potential applications in quantum computing.

In observing our model's success with "well-known" \(f\overline{f}\) systems, we find promising indications. These outcomes pave the way for an extension of our model to encompass $q\overline{q}$ pairs by choosing $\mathcal{S}(r)=-\alpha/r+kr$ under the influence of an external magnetic field.

\subsection{A conditionally exact solution for Cornell potential $\mathcal{S}(r)=-\alpha/r+kr$ and \(\mathcal{B}_{0} \neq 0\)}\label{sec3.2}

In this section, we will undertake analytical solutions of Eq. (\ref{WE}) for a $q\overline{q}$ pair subjected to the Cornell potential $\mathcal{S}(r)=-\alpha/r+kr$ under the influence of an external magnetic field. In this scenario, the wave equation can be explicitly expressed as follows:
\begin{equation}
\chi^{''}_{1}+\frac{2}{r}\chi^{'}_{1}+\left[\frac{\textcolor{black}{\tilde{\mathcal{E}}}^2-\left(2\tilde{m}-\frac{\alpha}{r}+kr\right)^2}{4}-\frac{4\left(s+\mathcal{B}r^2\right)^{2}}{r^2}\right]\chi_{1}=0. \label{3.1}
\end{equation}
By considering an ansatz that reads as $\chi_{1}(r)=\frac{1}{r}\chi(r)$, the wave equation can be expressed in the following form:
\begin{equation*}
\chi^{''}+\left[\frac{\textcolor{black}{\tilde{\mathcal{E}}}^2}{4}-\textcolor{black}{V^{2}_{eff}(r)}\right]\chi=0,
\end{equation*}
where
\begin{equation*}
\textcolor{black}{\tilde{\mathcal{E}}=\frac{\mathcal{E}}{\hbar c}},\quad \textcolor{black}{V_{eff}=\sqrt{\frac{\left(2\tilde{m}-\frac{\alpha}{r}+kr\right)^2}{4}+\frac{4\left(s+\mathcal{B}r^2\right)^{2}}{r^2}}}.
\end{equation*}
\textcolor{black}{Here, it is clear that \( \tilde{\mathcal{E}} \) has units of inverse length, indicating that \( V_{\text{eff}} \) also has units of inverse length. Therefore, \( \hbar c\, V_{\text{eff}} \) is in units of energy. Moreover, it is apparent that the parameter \( k \) has units of inverse length squared, while \( kr \) is measured in units of inverse length.} \\

The Eq. (\ref{3.1}) can be simplified as the following
\begin{equation*}
\chi^{''}_{1}(r)+\frac{2}{r}\chi^{'}_{1}(r)+\left(\frac{\alpha \tilde{m}}{r}-\frac{\tilde{S}^2}{r^2}-\tilde{\mathcal{B}}^2 r^2-k\tilde{m}r-\lambda\right)\chi_{1}(r)=0,
\end{equation*}
where
\begin{equation*}
\begin{split}
&\tilde{S}^2=4s^2+\frac{\alpha^2}{4},\quad \tilde{\mathcal{B}}^2=4\mathcal{B}^2+\frac{k^2}{4},\\
&\lambda=8\mathcal{B}s-\frac{\alpha k}{2}+\tilde{m}^2-\frac{\textcolor{black}{\tilde{\mathcal{E}}}^2}{4}.
\end{split}
\end{equation*}
This equation admits a radial solution in the form of 
\begin{equation}
    \chi_{1}(r)=\mathcal{N}r^\sigma\, \exp\left(-\frac{\mathcal{B}r^2}{2}-\frac{k\tilde{m}r}{2\mathcal{B}}\right)H_B(\tilde{\alpha},\tilde{\beta},\tilde{\gamma},\tilde{\delta};z),
\end{equation}
where
\begin{equation}
\begin{split}
&\sigma=-\frac{1}{2}+\frac{1}{2}\sqrt{1+4\tilde{S}^2},\quad \tilde{\alpha}=\sqrt{1+4\tilde{S}^2},\quad \tilde{\beta}=\frac{k\tilde{m}}{\tilde{\mathcal{B}}^{3/2}},\\
&\tilde{\gamma}=\frac{k^2\tilde{m}^2-4\tilde{\mathcal{B}}^2\lambda}{\tilde{\mathcal{B}}^3},\quad \tilde{\delta}=-\frac{2\alpha\tilde{m}}{\sqrt{\tilde{\mathcal{B}}}},\quad z=\sqrt{\tilde{\mathcal{B}}}r.
\end{split}
\end{equation}
One should note that we have intentionally ignored the second part of the radial function $\chi_{1}(r)$ because it is associated with $\sigma=-\frac{1}{2}-\frac{1}{2}\sqrt{1+4\tilde{S}^2}$, which leads to an infinite solution at $r=0$. Furthermore, the biconfluent Heun function/series $H_B(\tilde{\alpha},\tilde{\beta},\tilde{\gamma},\tilde{\delta};z)$ must be truncated to a polynomial of order $n_r+1\geq 1$ according to the detailed new method described in the Appendix of Mustafa \cite{omar-1} and utilized by Mustafa and colleagues \cite{omar-2,omar-3,omar-4}. Specifically, this method requires the truncation of the biconfluent Heun functions/series to a polynomial of order $n_r+1\geq 1$ when two conditions
\begin{equation}
\tilde{\gamma}=2(n_r+1)+\tilde{\alpha},\label{FC}
\end{equation} 
and
\begin{equation}
\tilde{\delta}=-\tilde{\beta}(2n_r+\tilde{\alpha}+3),\label{SC}
\end{equation}
are met. \textcolor{black}{The second condition makes it possible to achieve conditional exact solvability by establishing a parametric correlation, while the first provides the quantization guidelines for a specific number of $n_r$ quantum states}. As a result, the radial wave function, which is physically admissible, is now expressed as:
\begin{equation}
    \chi_{1}(r)=\mathcal{N}r^\sigma\,\exp\left(-\frac{\mathcal{B}r^2}{2}-\frac{k\tilde{m}r}{2\mathcal{B}}\right)\sum\limits_{j=0}^{n_r+1 }C_{j}\,y^{j},
\end{equation}
with $C_0=1$, 
\begin{equation}
    C_1=\frac{k\tilde{m}\sigma-\tilde{\mathcal{B}}P_2}{2\tilde{\mathcal{B}}(\sigma+1)}C_0, \quad P_2=\alpha \tilde{m}-\frac{k\tilde{m}}{\tilde{\mathcal{B}}},
\end{equation}
and  $\forall j\geq 0$. One can derive $C_j's$ using the three-term recursion relation
\begin{gather}
    C_{j+2}
    =\frac{C_{j+1}\left[P_2-\frac{k\tilde{m}}{\tilde{\mathcal{B}}}(j+\sigma+1)\right] 
    +C_{j}\left[P_1-2\tilde{\mathcal{B}}(j+\sigma)\right]}{\tilde{S}^2-(j+\sigma+2)(j+\sigma+3)},
\end{gather}
where
\begin{equation*}
   P_1=\frac{k^2\tilde{m}^2}{4\tilde{\mathcal{B}}^2}-\lambda-3 \tilde{\mathcal{B}}. 
\end{equation*}
At this juncture, readers are encouraged to consult the Appendix of \cite{omar-1} for additional insights into this matter. In the context of such conditional exact solvability, the condition outlined in Eq. (\ref{SC}) would readily produce results, as elaborated in detail (see \cite{omar-2,omar-3,omar-4}), 
\begin{equation}
k\rightarrow k_{\alpha}= \frac{4\alpha \mathcal{B}}{\sqrt{\left(\rho^{n_{r}}_{\alpha}\right)^2-\alpha^2}},\quad \rho^{n_{r}}_{\alpha}= \left(2n_{r}+3+\tilde{\alpha}\right).\label{second}
\end{equation}%
Substituting Eq. (\ref{second}) into Eq. (\ref{FC}), we derive the subsequent energy expression, as also detailed in \cite{omar-1}, expressed as
\begin{widetext}
\begin{equation}
\mathcal{E}_{ns}=\sqrt{4\tilde{m}^2c^2\hbar^2\left[1-\frac{\alpha^2}{\left(\rho^{n_{r}}_{\alpha}\right)^2}\right]+2\mathcal{B}c^2\hbar^2\sqrt{\frac{\left(\rho^{n_{r}}_{\alpha}\right)^2}{\left(\rho^{n_{r}}_{\alpha}\right)^2-\alpha^2}}\left[\rho^{n_{r}}_{\alpha}-1\right]-\frac{8\mathcal{B}\alpha^2c^2\hbar^2}{\sqrt{\left(\rho^{n_{r}}_{\alpha}\right)^2-\alpha^2}}+32\mathcal{B}c^2\hbar^2 s},\label{g-spectra}
\end{equation}
\end{widetext}
\textcolor{black}{for $q\overline{q}$ pair subject to Cornell-type Lorentz scalar potential} under the effects of an external uniform magnetic field. When $\mathcal{B}_{0}$ equals zero, this energy spectrum can be simplified to the following form
\begin{equation}
\mathcal{E}_{ns}=\pm 2m_{q}c^2\sqrt{1-\frac{\alpha^2}{\left(\rho^{n_{r}}_{\alpha}\right)^2}}\label{r-s-1}.
\end{equation}
We can derive binding energy levels, in principle, for quarkonium systems by employing the power series expansion method on Eq. (\ref{r-s-1})
\begin{widetext}
\begin{equation}
\begin{split}
&\mathcal{E}_{ns}\approx 2m_{q}c^2 \left\lbrace 1-\frac{8}{9}\frac{ \alpha^{2}_{s}}{\left(2n_{r}+3+\sqrt{16s^2+1}\right)^2} + \frac{32}{81}\frac{ \left(8n_r+12+3\sqrt{16s^2+1}\right)\,\alpha^{4}_{s}}{\left(2n_{r}+3+\sqrt{16s^2+1}\right)^4 \sqrt{16s^2+1}} +\mathcal{O}\left(\alpha^{6}_{s}\right) \right\rbrace\\
&\\
&\Rightarrow \mathcal{E}^{b}_{ns}\approx -\frac{16 m_{q}c^2\,\alpha^{2}_{s}}{9\left(2n_{r}+3+\sqrt{16s^2+1}\right)^2}\left[1-\frac{2}{9}\frac{\left(8n_{r}+12+3\sqrt{16s^2+1}\right)\,\alpha^{2}_{s}}{\left(2n_{r}+3+\sqrt{16s^2+1}\right)^2\sqrt{16s^2+1}}\right] \\
&\\
&\Rightarrow \mathcal{E}^{b}_{n0}\approx -\frac{4 m_{q}c^2\,\alpha^{2}_{s}}{9\left(n_{r}+2\right)^2}\left[1-\frac{1}{18}\frac{\left(8n_{r}+15\right)\,\alpha^{2}_{s}}{\left(n_{r}+2\right)^2}\right],\label{q-s-e}
\end{split}
\end{equation}
\end{widetext}
in the event that the binding energy is negative, through substitution \( \alpha \rightarrow \frac{4}{3} \alpha_{s} \), \textcolor{black}{where $\alpha_{s}$ is the strong coupling constant (in reality, $\alpha_{s}$ is not simply a constant but a running coupling parameter \cite{HF3}). This clarifies that the energy expressions (\ref{q-s-e}) solely encompass quadratic terms of the total spin (\( s^2 \)) of the resultant composite particle. Consequently, the spin quantum states corresponding to \( s=\pm 1 \) emerge as degenerate states in the absence of an external magnetic field. However, as depicted by the energy spectra (\ref{g-spectra}), the presence of a magnetic field induces a splitting of these states due to the presence of the additional term proportional to \( \mathcal{B}\,s \). Based on our findings (bearing in mind that the binding energy is negative), one can derive mass spectra, in principle, using \( \mathcal{E}_{ns} - |\mathcal{E}^{b}_{ns}| \). Additionally, transition energies can be calculated using \( \mathcal{E}^{b}_{ns} \) for quarkonium systems under various experimental conditions, such as when \( \mathcal{B}_{0} = 0 \) or \( \mathcal{B}_{0} \neq 0 \).} 

\textcolor{black}{Here, it is worth noting that we associated the parameter \( k \) with \( \alpha \) in the quasi-exact solution for the scenario where \( \mathcal{S}=kr-\alpha/r \). This means that one needs to use effective values of \( \alpha_s \) to fit experimental data. The result (\ref{r-s-1}) allow us to calculate the hyperfine splitting (HF), in principle. By taking \( m_{b}=4880 \) [MeV/c\(^2\)] \cite{mass} and using an effective value of \( \alpha_{s}=0.36 \) for bottomonium (\( b\overline{b} \)) in its ground state, our calculations yield \( \Delta \mathcal{E}_{HF} = \mathcal{E}_{01} - \mathcal{E}_{00}\sim 60 \) [MeV]. This finding is consistent with the HF value reported earlier in \cite{HF1, HF2} (see also \cite{HF3}) for \( b\overline{b} \). However, it is crucial to underline that we have taken \( \alpha_s \) as a constant, in principle.} \textcolor{black}{In fact, the \( \alpha_s \) is running coupling constant \( \alpha_s(Q^2) \) (or \( \alpha_s(r) \) ) in the context of quarkonium systems such as charmonium (\( c\overline{c} \)) and \( b\overline{b} \). \( \alpha_s(Q^2) \) is a crucial parameter that determines the strength of the strong interaction between the \( q\overline{q} \) pairs. It varies with the energy scale \( Q \), which, for these systems, is typically related to the masses of the quarks and their relative velocity. For \( c\overline{c} \) and \( b\overline{b} \), the values of \( \alpha_s(Q^2) \) can be derived using the leading-order formula (or next-to-leading order formulas) (for more details see \cite{HF3}) \( \alpha_s(Q^2) = \frac{4\pi}{\beta_0 \ln(Q^2/\Lambda_{\text{QCD}}^2)} \) where \( \beta_0 \) is a coefficient that depends on the number of quark flavors, and \( \Lambda_{\text{QCD}} \) is the QCD scale parameter. Additionally, in a $2+1$ dimensional vacuum, the properties of quarkonium systems may exhibit significant differences compared to the $3+1$ dimensional case because the reduced dimensionality affects the confining potential and the behavior of the strong interaction, leading to changes in the mass spectra, binding energies, and HF of these systems. In $2+1$ dimensions, the potential between the \( q\overline{q} \) pair can be modeled more effectively by other appropriate interaction potentials \cite{pot-1,pot-2}. Here, it is worth underlining that Eq. (\ref{WE}) can be useful to arrive at analytic results for specific systems under scrutiny and also for acquiring the results to fit experimental data. We defer detailed discussions on this subject to researchers specialized in this field.}

The energy spectrum (\ref{g-spectra}) also provides precise relativistic Landau levels (when $\alpha=0$ and equivalently $k_{\alpha}=0$), pertaining specifically to the singlet quantum state ($s=0$) of a general \(f\overline{f}\) pair, as extensively elaborated in \cite{g-epjc}
\begin{equation}
\begin{split}
&\mathcal{E}_{n}=\pm m_{t}c^2\sqrt{1+\frac{2\omega_{c}\hbar}{m_{t}c^2}\tilde{n}},\quad \omega_{c}=\frac{|e|\mathcal{B}_{0}}{mc},\\
&m_{t}=2m, \quad \tilde{n}=(n_r+1),\quad (n_r=0,1,2..),\label{RLL1}
\end{split}
\end{equation}
in which $\omega_{c}$ represents the relativistic cyclotron frequency. It leads to $\mathcal{E}\rightsquigarrow \pm m_{t}c^2$ for a system involving massive fermions and antifermions when $\mathcal{B}_{0}\rightsquigarrow0$. Moreover, by substituting $c$ with $\nu_{f}$ (where $\nu_{f}$ denotes the Fermi velocity), Eq. (\ref{RLL1}) facilitates the derivation of Landau levels for a Weyl pair ($m=0$) within monolayer graphene sheets
\begin{equation}
\mathcal{E}_{n}=\pm 2 \frac{\hbar \nu_{f} }{\ell_{\mathcal{B}}}\sqrt{n+1},\label{WP}
\end{equation}
where $\ell_{\mathcal{B}}$ is the magnetic length \cite{graphene,mag.len}. This outcome provides precise results for the excited states of a Weyl particle in a monolayer graphene sheet under an external magnetic field (see \cite{graphene}) and moreover there is no distinguishing feature to differentiate these modes from one another. Therefore, our findings suggest that the observation of Landau levels in monolayer graphene sheets containing Weyl particles inherently encompasses many-body effects, especially when subjected to an external magnetic field. 

\section{Summary and discussions}\label{sec4}

Grasping the dynamics of \textcolor{black}{$f\overline{f}$} pairs is pivotal in contemporary physics. These pairs, influenced by the \textcolor{black}{fundamental} force, engage in intricate interactions that are fundamental to unraveling the complexities of particle physics. This paper introduces a novel theoretical framework utilizing the fully-covariant one-time many-body Dirac equation with position-dependent mass setting to explore these systems. In order to authenticate our model, we verify its precision by aligning our findings meticulously with known energy eigenvalues for $f\overline{f}$ pairs regulated by the Coulomb potential, without relying on any approximations. Moreover, by integrating adaptations rooted in the Cornell potential, we formulate an \textcolor{black}{energy expression for $q\overline{q}$ systems under the influence of an external uniform magnetic field}. Our investigation rigorously scrutinizes the attributes of $f\overline{f}$ states, notably under the influence of magnetic field. We think that our distinctive model provides pivotal insights into the dynamics of $f\overline{f}$ pairs, encompassing their interactions and potential ramifications in particle physics. Furthermore, the versatility of our model transcends to diverse $f\overline{f}$ systems, spanning positronium, excitons in monolayer TMDCs, relativistic Landau levels for $f\overline{f}$ pairs, and Weyl pairs within monolayer graphene structures. 

To summarize, we have investigated the singlet-positronium model (where the orbital quantum number is not a good quantum number), disregarding the influence of external magnetic fields. This enabled us to ascertain the precise energy spectrum:
\begin{widetext}
\begin{equation}
\mathcal{E}_{n}\approx 2m_{e}c^2 \left\lbrace 1-\frac{\alpha^2}{8\,(n+1)^2}+\frac{(8n+7)\,\alpha^4}{128\,(n+1)^4}-\frac{(16n^3+72n^2+88n+33)\,\alpha^6}{1024\,(n+1)^6} \right\rbrace .\label{con-1}
\end{equation}
\end{widetext}
In this context, the initial component signifies the overall rest mass energy ($2m_{e}c^2$), while the term proportionate to $\alpha^2$ correlates to the renowned non-relativistic binding energy. For para-positronium in ground state, this binding energy is estimated as $\mathcal{E}^{b}_{0} \approx -m_{e}c^2\alpha^2/4 \sim -6.8$ eV \cite{positronium} assuming $m_{e}$ as the standard electron mass. The subsequent terms of higher order ($\alpha^4, \alpha^6$) encompass relativistic corrections. Hence, the energy carried by the annihilation photons (since para-positronium predominantly decays into two photons to uphold charge conjugation parity) can be computed as $\sim 2m_{e}c^2 - |\mathcal{E}^{b}|$. The binding energy of para-positronium is documented in Table \ref{tab:table1}. 

Subsequently, we expanded our findings to ascertain the precise energy spectra for excitons in monolayer TMDCs by substituting $\alpha$ with $\alpha/\epsilon_{eff}$, with $\epsilon_{eff}$ representing the effective dielectric constant of the surrounding medium. We offer an elaborate analysis of the binding energy of these excitons in Table \ref{tab:table2}, showcasing that our model adeptly elucidates the variances in observed exciton binding energies within identical monolayer TMDCs positioned on diverse substrates. 

\textcolor{black}{Next, we have obtained an analytic solution for $f\overline{f}$ pairs subject to Cornell-type Lorentz scalar potential} in the presence of an external magnetic field. This procedure entailed achieving a conditionally precise solution, extensively outlined in \cite{omar-1}, of the biconfluent Heun equation, culminating in a non-perturbative, closed-form energy expression:
\begin{widetext}
\begin{equation}
\mathcal{E}_{ns}=\sqrt{4\tilde{m}^2c^2\hbar^2\left[1-\frac{\alpha^2}{\left(\rho^{n_{r}}_{\alpha}\right)^2}\right]+2\mathcal{B}c^2\hbar^2\sqrt{\frac{\left(\rho^{n_{r}}_{\alpha}\right)^2}{\left(\rho^{n_{r}}_{\alpha}\right)^2-\alpha^2}}\left[\rho^{n_{r}}_{\alpha}-1\right]-\frac{8\mathcal{B}\alpha^2c^2\hbar^2}{\sqrt{\left(\rho^{n_{r}}_{\alpha}\right)^2-\alpha^2}}+32\mathcal{B}c^2\hbar^2 s}.\label{con-2}
\end{equation}
\end{widetext}
Notably, such an energy expression offers means to determine the binding energies and mass distributions for quarkonium systems, provided the binding energy is negative. It reveals that the energy expression consists solely of quadratic terms of the total spin (\( s^2 \)) of the resulting composite particle in the absence of an external magnetic field ($\mathcal{B}=0$), indicating that the spin quantum states corresponding to \( s=\pm1 \) are degenerate when $\mathcal{B}=0$. However, the introduction of a magnetic field leads to the splitting of these states due to the additional term proportional to \( \mathcal{B}s \). \textcolor{black}{In principle, from this outcome, considering realistic values of the running coupling constant $\alpha_{s}(Q^2)$, one can theoretically derive mass distributions, and the binding energy for quarkonium systems under diverse experimental conditions, such as $\mathcal{B}_{0}=0$ or $\mathcal{B}_{0}\neq0$. As an application, the results in Eq. (\ref{q-s-e}) can be used to calculate the HF by using an effective coupling value of \(\alpha_{s}\). For $b\overline{b}$ in its ground state, our calculations yield \(\Delta \mathcal{E}_{HF} = \mathcal{E}_{01} - \mathcal{E}_{00} \sim 60\) [MeV] by using \(\alpha_{s} = 0.36\). This result is consistent with the HF values previously reported in \cite{HF1, HF2} for \(b\overline{b}\) in its ground state. However, as emphasized before, $\alpha_s$ is not simply a constant, on the contrary, it is a running coupling parameter (see \cite{HF3}). Therefore, detailed analysis and discussions on the subject are best left to researchers who specialize in it.} 

Ultimately, we utilized this energy spectrum in the analysis of established $f\overline{f}$ configurations. It was illustrated that the energy spectrum (\ref{con-2}) accurately predicts relativistic Landau levels (when $\alpha=0$ and equivalently $k_{\alpha}=0$), specifically concerning the singlet quantum state ($s=0$) of a generic $f\overline{f}$ pair, as extensively detailed in \cite{g-epjc}:
\begin{equation}
\begin{split}
&\mathcal{E}_{n}=\pm m_{t}c^2\sqrt{1+\frac{2\omega_{c}\hbar}{m_{t}c^2}\tilde{n}},\quad \omega_{c}=\frac{|e|\mathcal{B}_{0}}{mc},\\
&m_{t}=2m, \quad \tilde{n}=(n_r+1),\quad (n_r =0,1,2..),\label{RLL}
\end{split}
\end{equation}
where $\omega_{c}$ is the relativistic cyclotron frequency. As a consequence, $\mathcal{E} \rightsquigarrow \pm m_{t}c^2$ arises for a configuration comprising massive fermions and antifermions when $\mathcal{B}_{0} \rightsquigarrow 0$. Additionally, with the substitution of $c$ by $\nu_{f}$ (where $\nu_{f}$ signifies the Fermi velocity), Eq. (\ref{RLL}) enables the computation of Landau levels for a Weyl pair ($m=0$) within monolayer graphene sheets:
\begin{equation}
\mathcal{E}_{n}=\pm 2 \frac{\hbar \nu_{f} }{\ell_{\mathcal{B}}}\sqrt{n_r+1},\label{WP}
\end{equation}
where $\ell_{\mathcal{B}}$ is the magnetic length \cite{graphene,mag.len}. This result yields accurate findings regarding the excited states of a Weyl particle within a monolayer graphene sheet subjected to an external magnetic field (refer to \cite{graphene}). Intriguingly, we discern no discernible feature to distinguish these modes from each other. This suggests that the observations (for further details, see  \cite{graphene,mag.len}) of the Landau levels of a Weyl particle within monolayer graphene sheets inherently encompass many-body effects, especially in the presence of an external magnetic field.

Finally, the current discussions do not encompass certain specialized systems and scenarios that warrant exploration in the near future, a direction we anticipate researchers will pursue. Specifically, in our examinations of positronium and excitons, we have not addressed the influence of magnetic fields, leaving this significant avenue open for further investigation. Moreover, the introduction of an external electric field has the potential to enrich all aspects of our research. Within the domain of the two-body problem, achieving such meticulous results may lay a robust groundwork for observing intriguing phenomena in each instance. Furthermore, investigating the impact of external electric fields on excitons within dielectric media holds promise for uncovering the captivating properties of Janus monolayers of TMDCs (for further elucidation, refer to \cite{janus}). To the best of our knowledge,  our methodical proposal discussed above has never been published elsewhere. The applicability of which may cover areas that are of crucial interest for condense matter physics, particle physics, and quantum gravity, to mention a few, including topological defects manifestly introduce by gravitational fields that are by products of different spacetime backgrounds \cite{td1,td2,td3,td4}. The confluent and biconfluent Heun functions are usually unavoidably feasible solutions for such spacetime backgrounds that are indulged with topological defects like, cosmic strings,  global monopoles, and domain walls. The current proposal offers a useful and handy approach that facilitates conditional exact solvability for a vast set of energy states that allows one to study and reliably elaborate on the effects of the gravitational fields on such systems. 

\begin{acknowledgments}
\textcolor{black}{The authors are grateful to the anonymous reviewers for their thorough reviews, valuable comments, and kind suggestions}.
\end{acknowledgments}



\end{document}